\documentclass[secnumarabic, superscriptaddress, amssymb, nobibnotes, aps,prl,twocolumn,longbibliography]{revtex4-1}

\usepackage{amsmath}
\usepackage{graphicx}
\usepackage{mathtools}
\usepackage{xcolor}
\usepackage{todonotes}
\usepackage{hyperref}
\usepackage{gensymb}
\usepackage{ulem}
\usepackage{multirow}

\begin{document}
\title{Influence of Bi alloying on GaAs valence band structure}

%authors here
\author{Joshua J.P. Cooper}
\affiliation{Department of Materials Science and Engineering, University of Michigan, Ann Arbor, MI 48109, USA}

\author{Jared W. Mitchell}
\affiliation{Department of Physics, University of Michigan, Ann Arbor, MI 48109, USA}

\author{Shane Smolenski}
\affiliation{Department of Physics, University of Michigan, Ann Arbor, MI 48109, USA}

\author{Ming Wen}
\affiliation{Department of Chemistry, University of Michigan, Ann Arbor, MI 48109, USA}

\author{Eoghan Downey}
\affiliation{Department of Physics, University of Michigan, Ann Arbor, MI 48109, USA}

\author{Chris Jozwiak}
\affiliation{Advanced Light Source, Lawrence Berkeley National Laboratory, Berkeley, CA 94720, USA}

\author{Aaron Bostwick}
\affiliation{Advanced Light Source, Lawrence Berkeley National Laboratory, Berkeley, CA 94720, USA}

\author{Eli Rotenberg}
\affiliation{Advanced Light Source, Lawrence Berkeley National Laboratory, Berkeley, CA 94720, USA}

\author{Kai Sun}
\affiliation{Department of Physics, University of Michigan, Ann Arbor, MI 48109, USA}

\author{Dominika Zgid}
\affiliation{Department of Physics, University of Michigan, Ann Arbor, MI 48109, USA}
\affiliation{Department of Chemistry, University of Michigan, Ann Arbor, MI 48109, USA}

\author{Na Hyun Jo}
\altaffiliation{nhjo@umich.edu}
\affiliation{Department of Physics, University of Michigan, Ann Arbor, MI 48109, USA}

\author{Rachel S. Goldman}
\altaffiliation{rsgold@umich.edu }
\affiliation{Department of Materials Science and Engineering, University of Michigan, Ann Arbor, MI 48109, USA}
\affiliation{Department of Physics, University of Michigan, Ann Arbor, MI 48109, USA}

\date{\today}

%%%%%%%%%%%%%%%%%%%%%%%%
% notation tools
\def\kill #1{\sout{#1}}
\def\add #1{\textcolor{blue}{#1}} 
\def\addred #1{\textcolor{red}{#1}} 
\newcommand{\tocite}[1]{$^\textbf{\textcolor{red}{CITE} #1}$ }
\newcommand{\postit}[1]{
    \begin{center}
        \fbox{
            \begin{minipage}{0.4\textwidth}
                \textit{#1}
            \end{minipage}
            }
    \end{center}
}
%%%%%%%%%%%%%%%%%%%%%%%%

\begin{abstract}
Bi alloying is predicted to transform GaAs from a semiconductor to a topological insulator or semi-metal. To date, studies of the GaAs$_{1-x}$Bi$_x$ alloy band structure have been limited, and the origins of Bi-induced enhancement of the spin-orbit splitting energy, $\Delta_\mathrm{SO}$, are unresolved. Here, we present high-resolution angle-resolved photoemission spectroscopy (ARPES) of droplet-free epitaxial GaAs$_{1-x}$Bi$_x$ films with $x_{\mathrm{Bi}}$ = 0.06. In addition to quantifying the Bi-induced shifts of the light-hole and heavy-hole valence bands, we probe the origins of the Bi-enhanced $\Delta_\mathrm{SO}$. Using exact-two-component density functional theory calculations, we identify the key role of Bi p-orbitals in the upward shift of the light-hole and heavy-hole bands that results in the Bi-enhanced $\Delta_\mathrm{SO}$.
 %support the Bi alloying-induced spin-orbit splitting enhancement, revealing that increased $\Delta_\mathrm{SO}$ occurs primarily through upward shifting of the light hole and heavy hole bands, with a smaller contribution due to the increased relativistic effect.
 %\color{blue} Theoretical DFT calculations reveal that the alloying Bi atoms are the main contributor to the enhanced spin-orbital coupling effect. \color{black}
\end{abstract}

\maketitle

\subsection{Introduction}
 
The discovery of topological materials compatible with compound semiconductors is a holy grail for quantum information science.
Of particular interest are the non-trivial topologies predicted for GaAs$_{1-x}$Bi$_x$ with tailored polytypes \cite{Fang2020} and/or sufficiently 
 %including topologically insulating behavior in GaAs$_{1-x}$Bi$_x$, with a 
high Bi compositions ($x_{\mathrm{Bi}}$ $\geq$ 0.19) \cite{Huang2014}.
 %and near-Dirac semimetal behavior for GaAs$_{1-x}$Bi$_x$ in the wurtzite phase and $x_{\mathrm{Bi}}$ = 0.50 \cite{Fang2020}. 
 %Growth of high Bi concentration GaAs$_{1-x}$Bi$_x$ is challenging, due to the tendency of Bi to segregate into droplets rather than incorporate into the film for typical GaAs growth conditions, with significant incorporation requiring relatively low growth temperatures \cite{tixier2003}. 
 %The suppression of Bi and Ga droplet formation during molecular beam epitaxy (MBE) growth requires careful consideration of growth temperature, flux ratios, and growth rate . Notably, growth using As$_4$, as an alternative to As$_2$, has been found to widen the window of V:III flux ratios available to avoid Bi droplet formation at high V:III ratios and Ga droplet formation at low V:III ratios.
Although compositions up to $x_{\mathrm{Bi}}$ = 0.22 have been reported \cite{lewis2012}, surface Ga and/or Bi droplet formation often limits the signal-to-noise ratio for surface-sensitive spectroscopies.  For droplet-free GaAs$_{1-x}$Bi$_x$ films \cite{lu2008, ptak2012, bastiman2012, vardar_2013, richards2014, field2016, bennarndt2016, carter2020}, electronic structure measurements
 %measurements of the direct band gap and split-off band to conduction band electronic transitions using photoluminescence spectroscopy \cite{ALBERI2018, ALGHAMDI2020, MANZOOR2022}, ellipsometry \cite{MANZOOR2022}, photocurrent \cite{LIU2021}, and photoreflectance spectroscopy \cite{ALBERI2015,BATOOL2012} 
have been limited to $x_{\mathrm{Bi}}$ $\leq$ 0.104 \cite{BATOOL2012, ALBERI2015, ALBERI2018, ALGHAMDI2020, LIU2021, MANZOOR2022}, with full momentum resolution limited to $x_{\mathrm{Bi}}$ $\leq$ 0.027 \cite{HONOLKA2019, OKABAYASHI2001}. 
 %Honolka et. al. studied the electronic structure of GaAs$_{1-x}$Bi$_x$ with $k$-resolved photoemission electron microscopy ($k$-PEEM) and ARPES, but the study was confined to low Bi composition of $x_{\mathrm{Bi}}$ = 0.027. 
 %While the energy- and momentum-resolved band structure of GaAs has been well documented by angle-resolved photoemission spectroscopy (ARPES) \cite{LARSEN1982,CHIANG1983,OLDE1990,CAI1992,KOBAYASHI2012}, systematic measurements of the effects of alloying \cite{HONOLKA2019,OKABAYASHI2001} and doping \cite{KOBAYASHI2012} in GaAs alloys have been limited. 
Meanwhile, Bi-induced enhancements of the spin-orbit splitting energy, $\Delta_\mathrm{SO}$, have been reported \cite{ALBERI2015,BATOOL2012}, with theoretical reports attributing the enhancements to upward shifting of the valence band maximum (VBM) by mixed Bi resonant states \cite{usman2011,kudrawiec2012,virkkala2013,joshya2014,bannow2016} or by Bi-induced relativistic spin-orbit coupling (SOC) \cite{zhang2005}.
 %It has been predicted, using a DFT-based charge patching method, that the increased splitting of the light hole (LH)/heavy hole (HH) and split-off (SO) bands is due to the relativistic spin-orbital coupling (SOC) effect of heavier and larger Bi atoms \cite{zhang2005}.
 %Others have interpreted the change in $\Delta$\textsubscript{SO} mainly as the result of the VBM shifting upwards, mainly due to mixed Bi resonant states, using parameterized valence band anti-crossing (VBAC) and tight binding (TB) models \cite{usman2011,kudrawiec2012,virkkala2013,joshya2014,bannow2016}.

In this work, we synthesize droplet-free epitaxial GaAs$_{1-x}$Bi$_x$ films with $x_{\mathrm{Bi}}$ up to 0.06, and use high-resolution angle-resolved photoemission spectroscopy (ARPES) to quantify Bi-induced shifts of the light-hole, heavy-hole, and SO valence bands. With $k\cdot p$ calculations of the band dispersion along $\Gamma$-K, we use exact-two-component density functional theory (x2C-DFT) to identify the key role of Bi p-orbitals in the upward shift of the light-hole and heavy-hole bands that yield Bi-enhanced $\Delta_\mathrm{SO}$. 
 %(GaAs$_{1-x}$Bi$_x$:Si), Si-doped GaAs (GaAs:Si), and undoped GaAs using MBE. We use high resolution ARPES to precisely measure the effects of Bi alloying and Si doping on the electronic structure of GaAs$_{1-x}$Bi$_x$ for $x_{\mathrm{Bi}}$ = 0.060. We discuss the change in energy of the LH, HH, and SO bands with the incorporation of Bi as well as compare detailed features in the curvature of the LH dispersion between GaAs, GaAs:Si, and GaAs$_{1-x}$Bi$_x$:Si. 
 %We also used theoretical methods to investigate the mechanisms of our experimental findings. We performed exact-two-component density functional theory (x2c-DFT) calculations on both pristine and Bi/Si-mixed GaAs lattices to explore the SOC effect. 
 %Furthermore, the $k\cdot p$ method was used to simulate the dispersion relations along the $\Gamma$-K high symmetry path. 
These new insights into the Bi-induced evolution of the GaAs valence band structure provide a critical step towards the development of III-V based topological insulators and semimetals.

\begin{figure*}
\includegraphics[width=\linewidth]{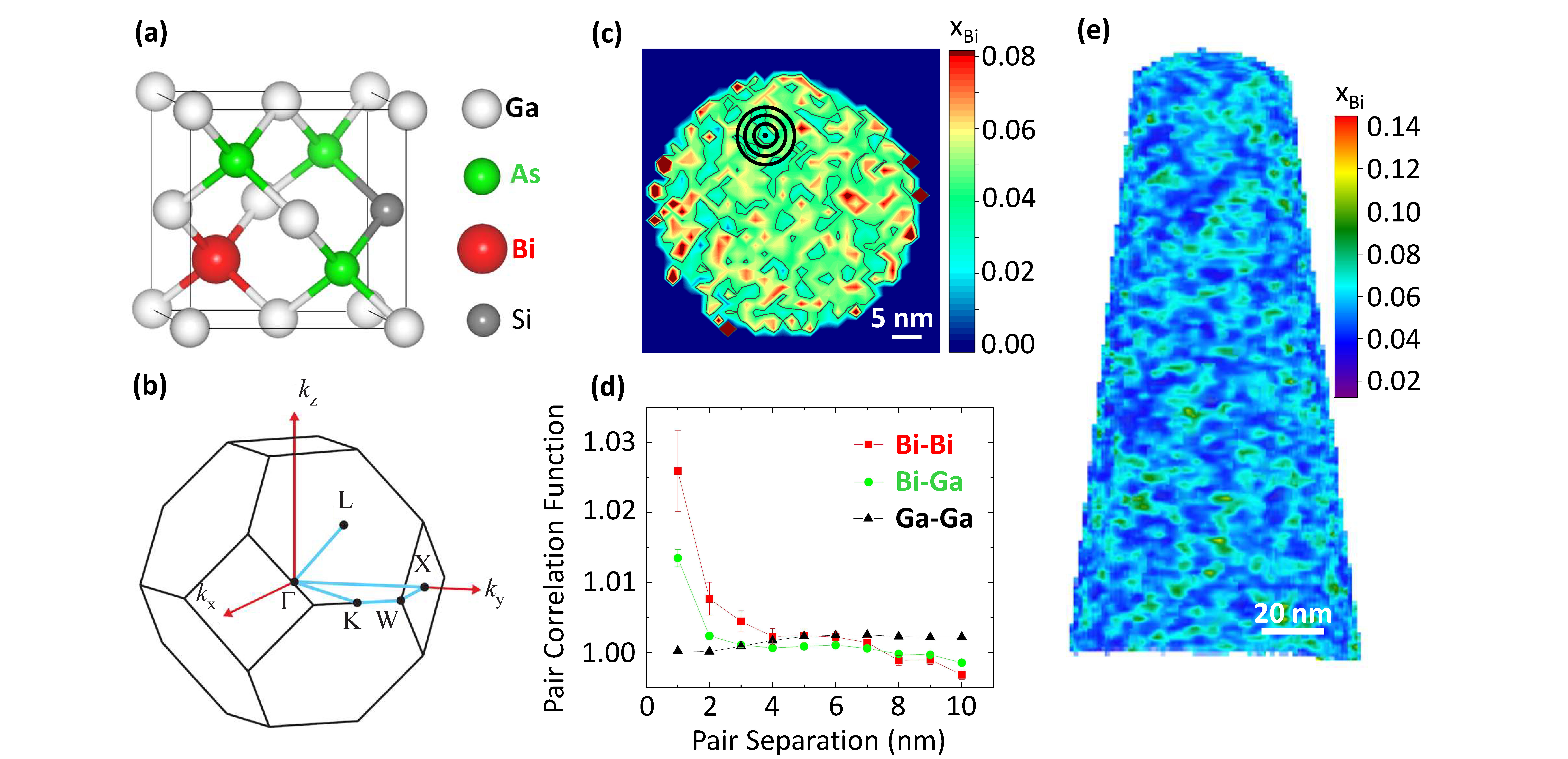}%
    \caption{GaAs$_{1-x}$Bi$_x$ atomic structure: (a) illustration of the GaAs unit cell containing Bi\textsubscript{As} (red) and Si\textsubscript{Ga} (gray). (b) GaAs Brillouin zone, with high-symmetry points labeled. Local $x_{\mathrm{Bi}}$ from local-electrode atom-probe tomography, shown as (c) a 2D contour plot created from a 1414 nm$^{3}$ cubic region-of-interest (bin size = 1.0 nm) and (e) a 3D rendering of the xz-cross section. (d) Pair correlation functions, C(r), vs pair separation for Bi–Bi (red), Bi–Ga (green), and Ga–Ga (black) pairs. Error bars are within the size of the data points for most values of C(r) \cite{mitchell_influence_2024}.
\label{fig:LEAPfig}}
\end{figure*}

\subsection{Methods}
%MBE Growth
For these investigations, a series of GaAs$_{1-x}$Bi$_x$ films were prepared by molecular-beam epitaxy (MBE) using $\geq$ 99.99999 $\%$ pure Ga and As, and $\geq$ 99.9999 $\%$ pure Bi and Si. The targeted layer thicknesses were determined using growth rate calibrations based upon reflection high-energy electron diffraction (RHEED) oscillations. Following oxide desorption and growth of an initial GaAs buffer layer at 580 \textdegree C, the substrate temperature was lowered to 500 \textdegree C and the sample was annealed for 5-10 minutes to achieve a flat buffer. In some cases, the substrate temperature was subsequently lowered to 340 \textdegree C for the growth of GaAs$_{1-x}$Bi$_x$ layers. 

For local-electrode atom probe tomography (LEAP) studies of the local Bi composition, a semi-insulating GaAs substrate was used for growth of a 500 nm-thick undoped GaAs buffer followed by an 500 nm-thick undoped GaAs$_{1-x}$Bi$_x$ layer. Following epitaxy, conical-shaped LEAP specimens were prepared by standard lift-out procedures and loaded into a Cameca LEAP 5000HR.  

 %MOVE TO SUPPLEMENTAL? For the n+ GaAs substrates and Si-doped GaAs buffers, the targeted free carrer concentrations were 2$\times$10$^{18}$ cm$^{-3}$ and 5$\times$10$^{17}$ cm$^{-3}$, respectively. For all films, the growth rate was 1.0 $\mu$m/hr. For the GaAs and GaAs$_{1-x}$Bi$_x$ layers, As$_4$/Ga beam equivalent pressure ratio (BEP) ratios of 25 and 19 were utilized. For the GaAs$_{1-x}$Bi$_x$ layers, the Bi BEP was 8.0$\times$10$^{-8}$ Torr. For all layers, the surface reconstruction was monitored in-situ with RHEED.  For arsenic capping, the source shutters were closed and the substrate temperature lowered to 60 \textdegree C prior to reopening the As$_4$ shutter. Following the disappearance of the RHEED pattern at $\sim$ 2 minutes, As$_4$ exposure was continued for an additional 60 minutes, yielding an $\sim$ 60 nm thick arsenic cap. 

\begin{figure*}
\includegraphics[width=\linewidth]{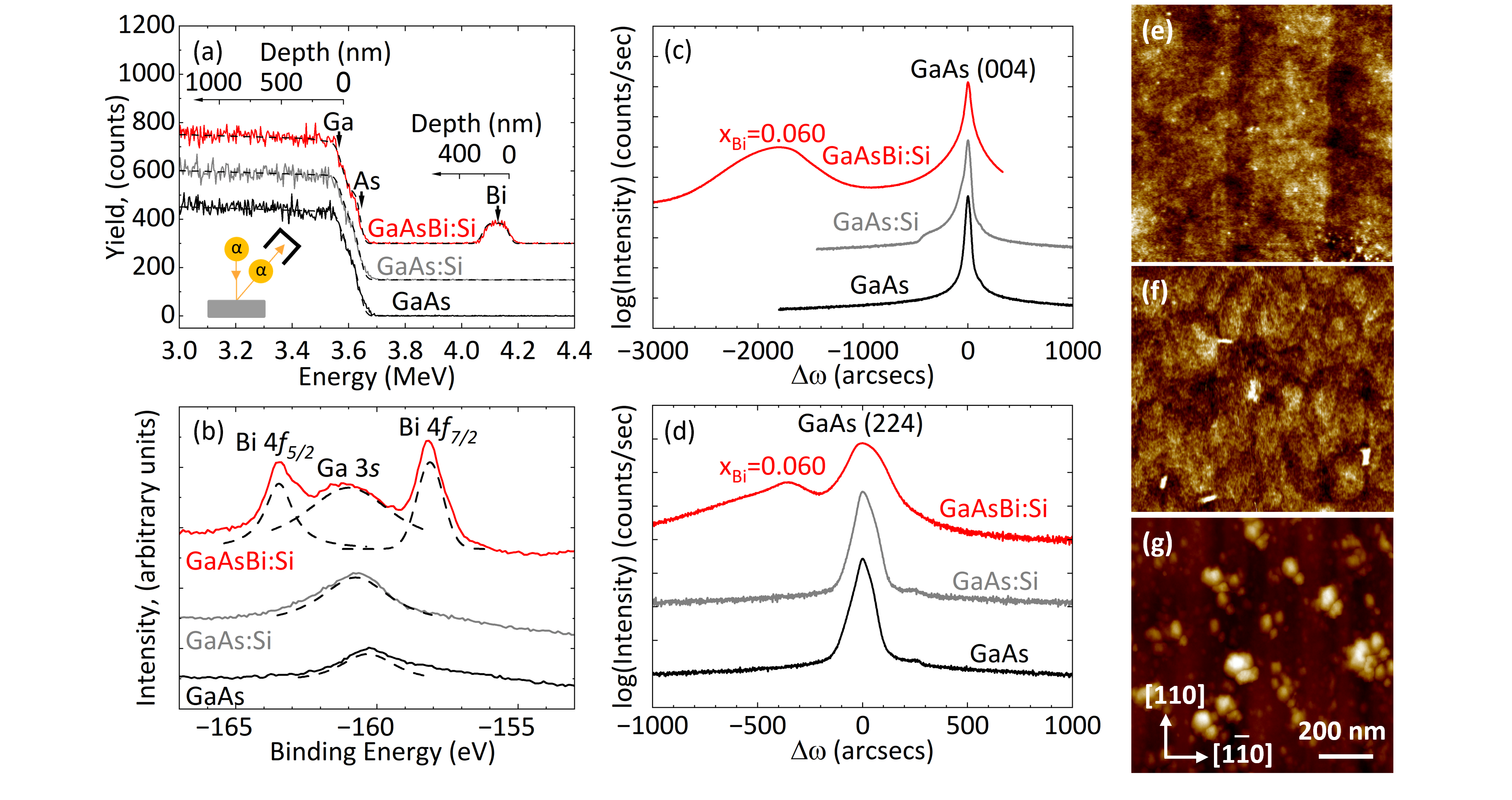}%
        \caption{GaAs$_{1-x}$Bi$_x$ film composition and morphology: (a) Rutherford backscattering spectrometry (RBS) yield versus backscattered particle energy (and depth) for  GaAs\textsubscript{1-x}Bi\textsubscript{x}:Si (red), GaAs:Si (gray), and GaAs (black) films. SIMNRA fitting of the GaAs$_{1-x}$Bi$_x$:Si spectrum yields an average Bi composition of $x_{\mathrm{Bi}}$ = 0.060 and layer thickness of 210 nm for the GaAs$_{1-x}$Bi$_x$:Si film. (b) Normalized XPS core level spectra for GaAs (black), GaAs:Si (gray), and GaAs\textsubscript{1-x}Bi\textsubscript{x}:Si (red) centered around the energies corresponding to the Bi 4\textit{f} and Ga 3\textit{s} core levels, with Voigt fits as dashed lines. 
        %Intensities have been scaled to match peak heights across samples.
        (c-d) high-resolution x-ray rocking curves (XRC), consisting of diffraction intensity vs. $\Delta\omega$ about the (c) (004) and (d) (224) GaAs for GaAs (black), GaAs:Si (gray), and GaAs$_{1-x}$Bi$_x$:Si (red) films. Analysis of the $\Delta\omega_{(004)}$ and $\Delta\omega_{(224)}$ data reveals a residual in-plane compressive strain of 0.62 $\%$ for the GaAs$_{1-x}$Bi$_x$:Si film. (e-g) atomic force microscopy (AFM) images for (e) GaAs$_{1-x}$Bi$_x$:Si, (f) GaAs:Si, and (g) GaAs. The color-scale ranges displayed are (e) 2.0 nm, (f) 1.5 nm, and (g) 23 nm.
\label{XRD_XPS_RBS}}
\end{figure*}
 
\begin{figure*} [ht]
\includegraphics[width=6.9 in]{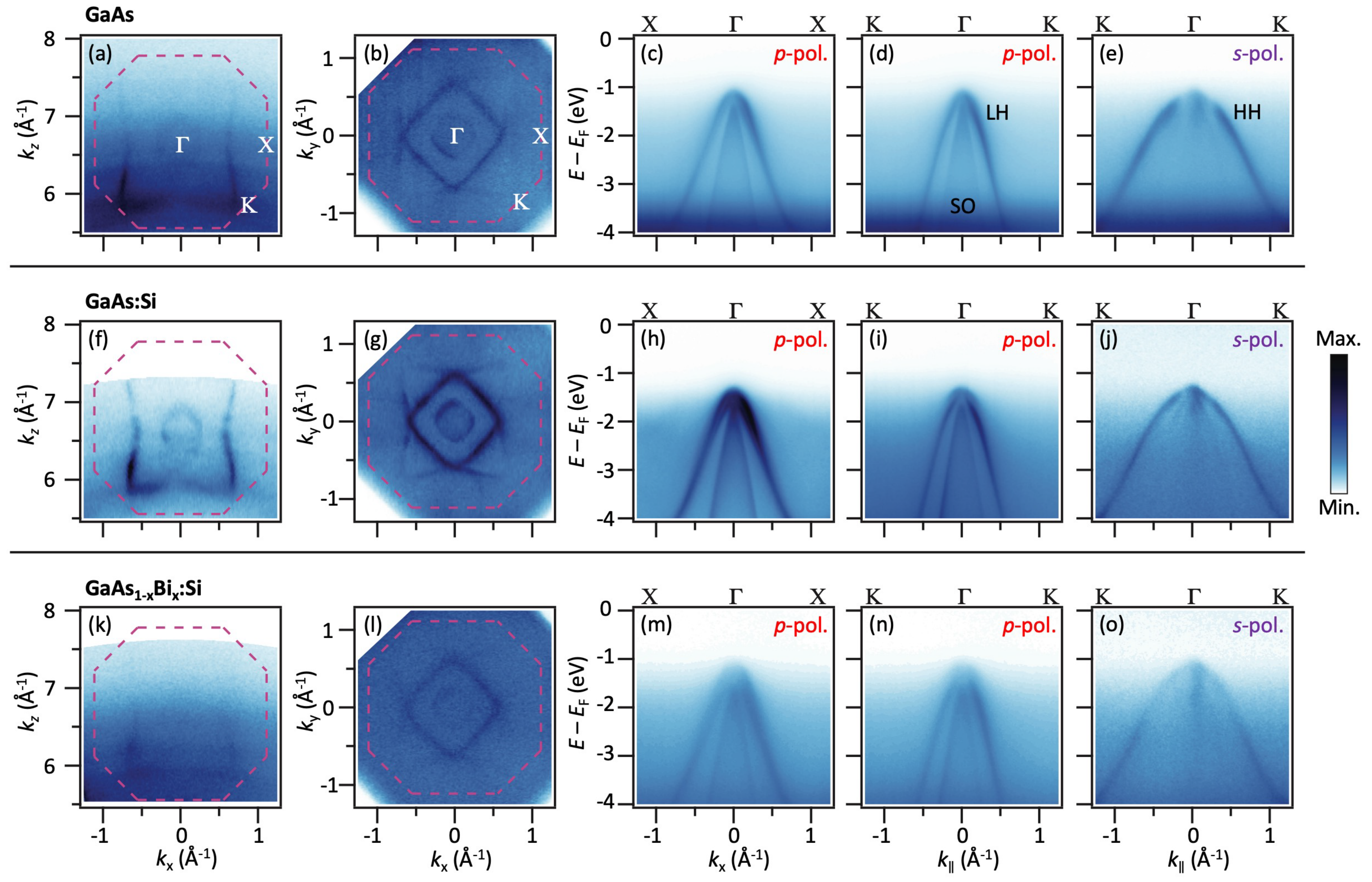}%
    \caption{Comparison of electronic structures for GaAs, GaAs:Si, and GaAs\textsubscript{1-x}Bi\textsubscript{x}:Si. (a) Out-of-plane and (b) in-plane constant energy contours at a binding energy of 3.5\,eV for GaAs taken with p-polarized light. The pinked dashed lines depict the first Brillouin zone with high symmetry points labeled. (c)-(e) Band dispersion of GaAs along the high symmetry lines (c) $\Gamma$\,-\,X with p-polarized light, (d) $\Gamma$\,-\,K with p-polarized light, and (e) $\Gamma$\,-\,K with s-polarized light. (f)\,-\,(j) and (k)\,-\,(o) are the same as (a)\,-\,(e) except for GaAs:Si and GaAs\textsubscript{1-x}Bi\textsubscript{x}:Si. 
\label{fig:Full}}
\end{figure*}

%ARPES/XPS Measurements
For ARPES studies, an n+ GaAs substrate was used for separate growths of a 500 nm-thick undoped GaAs buffer, a 500 nm-thick Si-doped GaAs buffer, and a 200 nm-thick Si-doped GaAs buffer followed by a 200 nm-thick Si-doped GaAs$_{1-x}$Bi$_x$ layer; the final epilayers were capped with arsenic. ARPES and x-ray photoelectron spectroscopy (XPS) were performed at Beamline 7.0.2 (MAESTRO) of the Advanced Light Source using a R4000 spectrometer with deflectors that enable stationary measurements of the entire Brillouin zone (shown in Fig. \ref{fig:LEAPfig} (b)). Prior to ARPES and XPS measurements, the arsenic cap was removed from each sample by annealing at 350\,\degree C for $\sim$45 minutes. The ARPES and XPS measurements were performed in a chamber with base pressure $\leq$ $5\times10^{-11}$ Torr, with measurement temperatures of $\sim$\,50\,K for GaAs:Si and $\sim$\,80\,K for GaAs and GaAs\textsubscript{1-x}Bi\textsubscript{x}:Si. 

Following arsenic-decapping, the spatially-averaged Bi compositions were determined using Rutherford backscattering spectrometry (RBS), 
 %in a 1.7 MV General Ionex tandem ion accelerator using 4.5 MeV $\alpha$ particles
the surface morphology was probed using tapping-mode atomic-force microscopy (AFM), and the residual strain was quantified using an analysis of high resolution x-ray rocking curves (XRC). 

For computational studies, Bi alloying and Si doping were modeled as Bi substituting for As (Bi$_{\mathrm{As}}$) and Si substituting for Ga (Si$_{\mathrm{Ga}}$), respectively, as illustrated in Fig. \ref{fig:LEAPfig} (a). 
 %To construct Ga$_8$As$_7$Bi and Ga$_7$SiAs$_8$ lattices, 
The band structures of GaAs, GaAs:Si, and GaAs$_{1-x}$Bi$_x$ were computed for $2\times 2\times 2$ GaAs supercells \cite{ku_synthesis_1968}, using the all-electron x2c-SVPall Gaussian type orbital basis set and PBE functionals \cite{perdew_generalized_1996,pollak_segmented_2017}, within the \texttt{pyscf} package~\cite{sun_gaussian_2017,sun_pyscf_2018,sun_recent_2020}. 
Although lattice strain and Bi-Bi interactions have been explored in earlier computational studies\cite{virkkala2013,bannow2016}, these effects are not considered in the x2C-DFT calculations, due to the large supercell size required.
 %Considerations of Bi-Bi interaction and the effect of lattice strain, as explored in previous studies \cite{virkkala2013,bannow2016}, are not included due to the large system size required to model these effects using x2C-DFT computations. 
To compute the $\Gamma$-K band dispersion (Fig. \ref{fig:LEAPfig} (b)), we constructed a $k\cdot p$ Hamiltonian for the HH, LH, and SO bands near the $\Gamma$ point. Assuming the $T_d$ space group and considering spin-orbit coupling, the fitting parameters were determined via fitting the computed LH band to the LH dispersion measured by ARPES.

 %supplemental? We used a $2\times 2\times 2$ \textit{k}-point sampling for Ga$_8$As$_8$, Ga$_8$As$_7$Bi, and Ga$_7$SiAs$_8$. The Fermi level was approximated as the HOMO-LUMO average energy (mid-gap approximation). The lattice parameters and atomic coordinates used for calculations can be found in the Supplemental Material. The projected density of state (PDOS) is calculated by summing over all $k$-points with a broadening factor of 0.05 eV. To keep the intensity of the Ga and As PDOS curves about the level of the Bi impurities for clear comparison, we calculated all of the PDOS curves using only one atom of each kind in the cell.

\subsection{GaAs$_{1-x}$Bi$_x$ film composition and morphology}

For the GaAs$_{1-x}$Bi$_x$ layers, the local Bi concentration profiles from LEAP data sets are shown in the 2D contour plot (Fig.~\ref{fig:LEAPfig} (c)) and volume rendering (Fig.~\ref{fig:LEAPfig}(e)). To quantify alloy disorder, we use a pair correlation function, C(r) = $\rho_{exp}(r)/\overline\rho(r)$, where $\overline\rho(r)$ is the average density of atomic species within each annular bin \cite{mitchell_influence_2024}. To determine $\rho_{exp}(r)$, the locations and separations between pairs of Ga and Bi atoms were determined from regions-of-interest in LEAP reconstructions spanning volumes \textgreater 1000 nm$^{3}$. Pairs of atomic species were binned in 1 nm intervals, yielding the number of experimental pairs, N$_{exp}$, in shells of multiples of 1 nm about a central atom up to r$_{max}$, as illustrated in Fig.~\ref{fig:LEAPfig} (c). Finally, $\rho_{exp}(r)$ is defined as N$_{exp}$ divided by the volume of each annual bin, with error bars determined by counting statistics. In Fig.~\ref{fig:LEAPfig} (d), C(r) vs. pair separation (in nm) are shown for Bi-Bi (red), Bi-Ga (green), and Ga-Ga (black) pairs.  For the Ga-Ga pairs,  C(r) values are within 0.5\% of unity, suggesting that Ga is randomly distributed in the layers.  Meanwhile, for both the Bi-Bi and Bi-Ga first nearest pairs, C(r) slightly exceeds unity, revealing Bi clustering in the GaAs$_{1-x}$Bi$_x$ layers.

As shown in Fig.~\ref{XRD_XPS_RBS}(a), the random RBS yields vs. backscattered particle energy (and depth) are shown the GaAs (black), GaAs:Si (grey), and GaAs$_{1-x}$Bi$_x$:Si (red) films, with the energies of the Ga, As, and Bi edges labeled in the plot. RBS data are overlaid with SIMNRA fitted spectra assuming a uniform Bi depth profile, shown as solid and dashed lines, respectively. For the GaAs (black), GaAs:Si (grey), and GaAs$_{1-x}$Bi$_x$:Si (red) films, the Ga and As edges are apparent. In addition, for the GaAs$_{1-x}$Bi$_x$:Si film, a distinct Bi peak is apparent at 4.15 MeV, confirming the incorporation of Bi into the film. SIMNRA fitting of the GaAs$_{1-x}$Bi$_x$:Si RBS yield indicates an average Bi composition of $x_{\mathrm{Bi}}$ = 0.060 and layer thickness of 210 nm.

The corresponding XPS core level spectra collected from GaAs (black), GaAs:Si (grey), and GaAs$_{1-x}$Bi$_x$:Si (red) surfaces are shown in Fig.~\ref{XRD_XPS_RBS}(b), with binding energies of the Ga 3\textit{s} and Bi 4\textit{f} core levels labeled in the plot. XPS data are overlaid with Voigt fits, shown in solid and dashed lines, respectively.  For the GaAs (black), GaAs:Si (grey), and GaAs$_{1-x}$Bi$_x$:Si (red) films, peaks at -160.7 $\pm$ 0.4 eV, corresponding to the Ga 3s core level, are clearly resolved. In addition, for the GaAs$_{1-x}$Bi$_{x}$:Si (red) film, additional peaks are apparent at -163.5 eV and -158 eV, corresponding to the Bi 4f$_{7/2}$ and Bi 4f$_{5/2}$ core levels, respectively. It is interesting to note that the presence of Bi in the GaAs$_{1-x}$Bi$_{x}$:Si film shifts the Ga 3\textit{s} core level to a higher binding energy, presumably due to the Bi-induced change in the local chemical environment of Ga ~\cite{STEVIE2020}.

Figure~\ref{XRD_XPS_RBS} (c-d) presents high-resolution XRC data, namely the diffraction intensity vs. $\Delta\omega$ about (c) (004) and (d) (224) GaAs for the GaAs (black), GaAs:Si (grey), and GaAs$_{1-x}$Bi$_{x}$:Si (red) films, with the GaAs substrate peak set to $\Delta\omega$ = 0 arcsec. For the GaAs (black) and GaAs:Si (grey) films, additional peaks are not apparent. On the other hand, for the GaAs$_{1-x}$Bi$_{x}$:Si (red) films, an additional peak is apparent at $\Delta\omega_{(004)}$ = -1814 arcsec and $\Delta\omega_{(224)}$ = -351 arcseconds; analysis of the $\Delta\omega_{(004)}$ and $\Delta\omega_{(224)}$ data reveals a residual in-plane compressive strain of 0.62 $\%$.  
 
The corresponding AFM micrographs in Fig.~\ref{XRD_XPS_RBS}(e-g) reveal droplet-free (\textless  $3.0 \times10^6$ cm$^{-2}$) surfaces; the color-scale ranges displayed are \textless 2.0 nm (GaAs and GaAs:Si) and 23 nm (GaAs$_{1-x}$Bi$_x$:Si). It appears that the As cap was partially-removed from the GaAs surface, and fully-removed from the GaAs:Si and GaAs$_{1-x}$Bi$_x$:Si surfaces. For the the GaAs$_{1-x}$Bi$_x$:Si surface (Fig.~\ref{XRD_XPS_RBS}(g)), $\sim$ 2 nm high terraces with a lateral period of $\sim$ 300 nm, may be related to step flow growth promoted by the Bi flux.

 %Importantly, Ga and As peaks are intense and clearly resolved at the expected energies in all three compositions while the Bi 4\textcolor{purple}{\textit{f}} peaks are present only in the Bi-doped films, confirming the expected chemical compositions. Si peaks are not resolved in either the GaAs:Si or the GaAs\textsubscript{1-x}Bi\textsubscript{x}:Si films as the concentration is below the detection limit. 

\subsection{Influence of Bi on GaAs valence bands}

To understand the effects of Bi alloying on the GaAs valence band dispersion in the vicinity of the $\Gamma$ point, we now consider ARPES data in conjunction with $k\cdot p$ theory and x2C-DFT.  Following a comparison of the valence band dispersions for GaAs, GaAs:Si, and GaAs$_{1-x}$Bi$_x$:Si films, we discuss the origins of the Bi-induced enhancement of $\Delta$\textsubscript{SO}.  

For undoped GaAs, the iso-energy plots in Figs.~\ref{fig:Full} (a) and (b) reveal the expected symmetry of the face-centered cubic GaAs lattice, enabling identification of the high symmetry directions. The ARPES valence band dispersion (Figs.~\ref{fig:Full} (c)-(e)) along the $\Gamma$-X and $\Gamma$-K reveals three parabolic hole bands, namely the heavy hole (HH), light hole (LH), and split-off (SO) bands, consistent with earlier theoretical and experimental reports~\cite{KOBAYASHI2012,LUO2009,usman2011,kudrawiec2012,virkkala2013,joshya2014,bannow2016}. Along the $\Gamma$-K direction, the LH and SO bands are apparent using \textit{p}-polarized incident light (Fig.~\ref{fig:Full} (d)), while the HH band is observed using \textit{s}-polarized light (Fig.~\ref{fig:Full} (e)). For both the $\Gamma$-X and $\Gamma$-K directions, the extracted HH and LH effective masses in the vicinity of the $\Gamma$ point are in reasonable agreement with literature values. However, for the SO band, the effective mass is $\sim$ 50 $\%$ of the reported value, possibly due to our limited averaging over two k-space directions, as discussed in the Supplementary Material \cite{blakemore1982}. 
 %\textcolor{orange}{The effective masses for each hole band along the $\Gamma$-K line are observed directly and listed in Table~\ref{table:EffectiveMasses}; the observed mass of the LH band is in good agreement with literature values while $m^*_{\mathrm{HH}}$ is roughly twice as large and $m^*_{\mathrm{SO}}$ roughly half as large as their literature values.}
At $\Gamma$, the LH and HH bands meet to form the valence band maximum (VBM) at $\sim$ 1.06\,eV below the Fermi level (\textit{E}\textsubscript{F}).  Meanwhile, the maximum of the SO band at $\Gamma$ is $\sim$  1.38\,eV below \textit{E}\textsubscript{F}. Thus, the energy difference between the LH/HH bands and the SO band at $\Gamma$, termed the spin-orbit splitting, ($\Delta$\textsubscript{SO}), is $\sim$ 320\,meV, in good agreement with earlier reports~\cite{LIU2021}.

We now consider the ARPES valence band dispersion for GaAs:Si (Figs.~\ref{fig:Full} (f)-(j)). For GaAs:Si, the HH, LH, and SO band dispersions are similar to those of pure GaAs with slight variations in the effective masses, as shown in the Supplemental Material. However, all GaAs:Si bands are shifted to lower energies in comparison to those of pure GaAs.  LH/HH band energies are reduced by 290\,meV, resulting in the VBM for GaAs:Si at 1.35\,eV below \textit{E}\textsubscript{F}. Similarly, the SO band energy is reduced by 300\,meV; thus, Si doping apparently has an insignificant effect on $\Delta$\textsubscript{SO}. The rigid downward shift of the valence bands in GaAs:Si is direct evidence of increased n-type doping induced by Si incorporation in our GaAs:Si film.

 \begin{figure}
    	\includegraphics[width=3.2in]{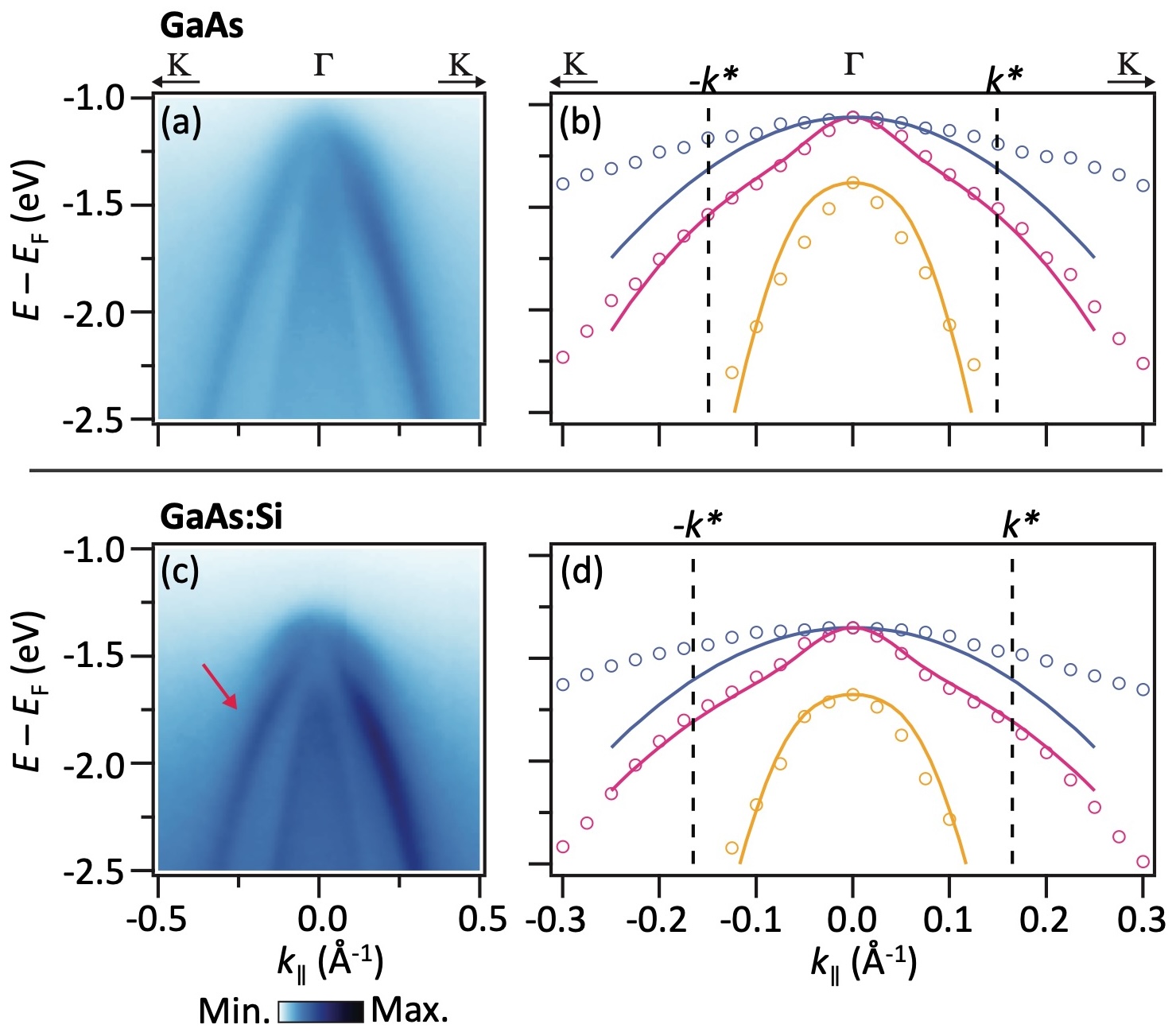}%
    	\caption{Effects of Si Doping on the Electronic Structure of GaAs. (a) Band dispersion taken with p-polarized light 
        %and (b) corresponding second derivative taken along the momentum axis 
        of GaAs along the $\Gamma$\,-\,K high symmetry line. (b) Band positions found through the energy-dispersion curves (EDCs) (markers) and the corresponding $k \cdot p$ fits (solid lines). The dashed lines denote $k*$, the limit at which 1st order $k \cdot p$ theory no longer is accurate. (c), (d) Same as (a),(b) but for GaAs:Si. The enhanced n doping of GaAs:Si is evident through the downward shift in energy of the bands relative to GaAs. The red arrow in (c) highlights a kink in the LH band dispersion of GaAs:Si that is absent in GaAs.
    \label{fig:Kink_Full}}
\end{figure}

 \begin{figure}
    	\includegraphics[width=3.2in]{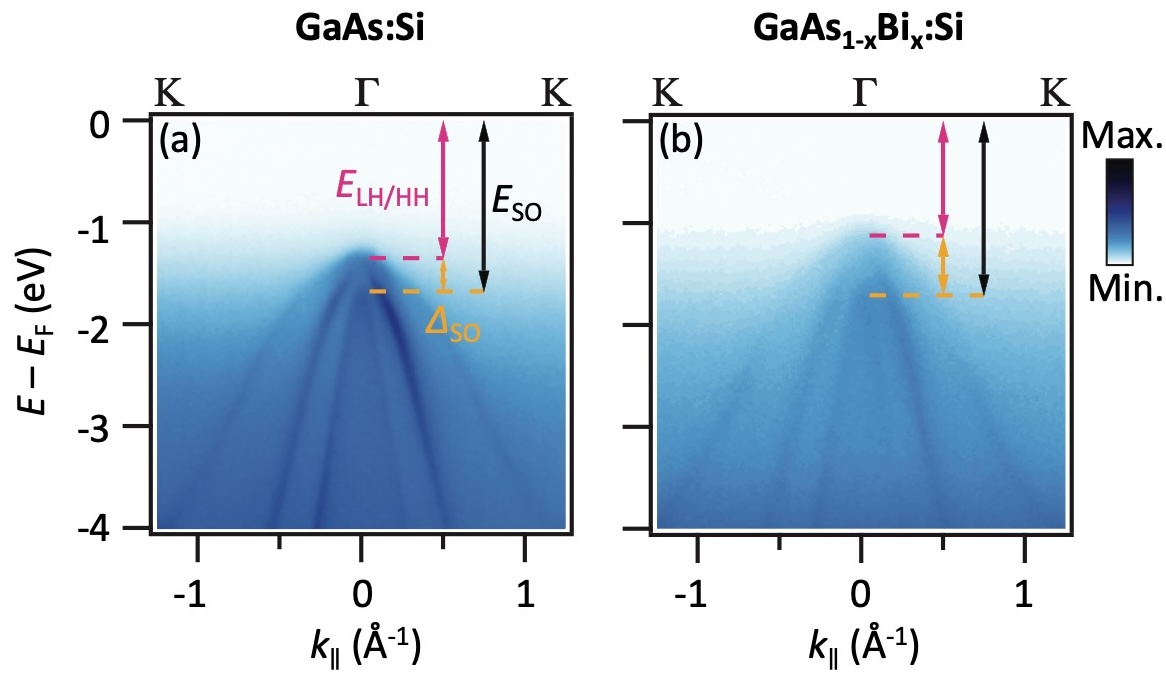}%
    	\caption{Enhanced Spin-Orbit Splitting in GaAs\textsubscript{1-x}Bi\textsubscript{x}:Si. (a) Energy dispersion and (b) the corresponding second derivative taken along the momentum axis of GaAs:Si along the $\Gamma$\,-\,K high symmetry line. (c)\,-\,(d) Same as (a)\,-\,(b), respectively, for GaAs\textsubscript{1-x}Bi\textsubscript{x}:Si. The pink and orange dashed lines show the LH/HH and SO band maxima, labeled as E\textsubscript{LH} and E\textsubscript{SO}, respectively. The difference between these band positions is the spin-orbit splitting, labeled as $\Delta$\textsubscript{SO}, which is enhanced in GaAs\textsubscript{1-x}Bi\textsubscript{x}:Si. All plots are a summation of the dispersion measured with p- and s-polarized light.
    \label{fig:Gaps}}
\end{figure}

\begin{figure}[ht]
    \centering
    \includegraphics[width=\linewidth]{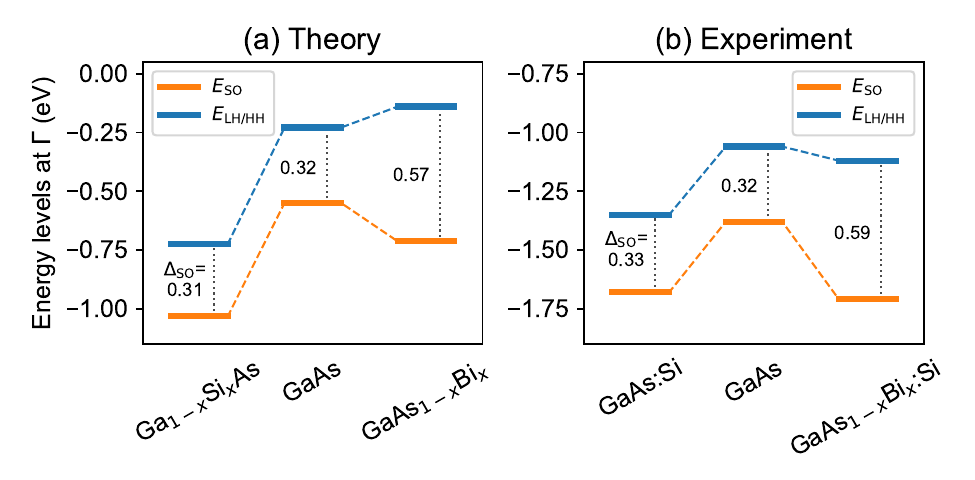}
    \caption{Energy levels for the GaAs light hole (LH) and split-off (SO) bands at $\Gamma$: (a) computed with x2c-DFT ($x=0.125$), and (b) measured using ARPES, as shown in FIG. \ref{fig:Gaps} ($x=0.06$). Both calculations and measurements reveal Bi-induced enhancements of the $\Delta_{\mathrm{SO}}$.} 
    \label{fig:SO-LH_split}
\end{figure}

\begin{figure}[ht]
    \centering
    \includegraphics[width=0.70\linewidth]{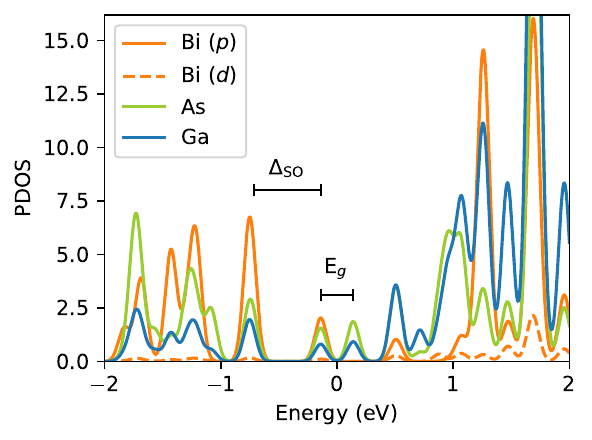}
    \caption{Atomic contributions to the GaAs$_{1-x}$Bi$_x$ ($x=0.125$) projected density of states (PDOS), computed with x2c-DFT, using a broadening factor of 0.05 eV. The plot includes Ga, As, and Bi $p$- and $d$-orbital contributions to the PDOS, along with the computed bandgap energy, E$_g$ and the spin-orbit splitting energy, $\Delta_{SO}$. It is apparent that $E_\mathrm{LH/HH}$ and $E_\mathrm{SO}$ consist of mixtures of Bi $p$ states and GaAs valence states. To ensure similar intensities of the computed atomic contributions to the PDOS, each unit cell contains a single atom. } 
    \label{fig:PDOS}
\end{figure}

 %Next, examining the shape of the dispersion of GaAs:Si more closely, we observe a slight kink in the LH band at $k~\approx~0.175$\,Å that is not present in the GaAs LH band (Fig.~\ref{fig:Kink_Full}). We analyze this more closely by examining how the group velocity \textit{v}\textsubscript{g} changes across this feature. In GaAs, \textit{v}\textsubscript{g} for $k>k_{\mathrm{kink}}$ is 30\% greater than that for $k<k_{\mathrm{kink}}$. This compares with GaAs:Si where \textit{v}\textsubscript{g} is 43\% greater for $k>k_{\mathrm{kink}}$ than $k<k_{\mathrm{kink}}$. To understand the origin of this feature and its effects on the dispersion near $\Gamma$, we employed a six-band, first-order $k\cdot p$ theory which we fit to our ARPES data for both GaAs and GaAs:Si near $\Gamma$ (Fig.~\ref{fig:Kink_Full} (c), (f)). From this fitting, we highlight two key observations. First, the $k\cdot p$ theory fits are in good agreement with our experimental data for $k<k^*$ where  $k^*$ is the outer boundary that $k\cdot p$ theory holds. Notably, the kink, which is found at $|k|\approx0.175$\,Å, falls beyond $k^*$, suggesting that it arises from higher order effects. More involved computational methods will be necessary to determine the exact origin of this feature. Next, all fitting parameters are nearly identical for both GaAs and GaAs:Si. Both of these insights from $k\cdot p$ theory suggest that the electronic near $\Gamma$ and, relatedly, $\Delta$\textsubscript{SO} are minimally affected by Si incorporation. 

We now consider the impact of Si doping on the GaAs valence band dispersion near $\Gamma$. 
 %A comparison of $k\cdot p$ theory fits to the ARPES data for GaAs and GaAs:Si (Fig.~\ref{fig:Kink_Full} (b), (d)). 
For $k$ values up to 
 %the outer boundary for $k\cdot p$ theory, 
$k^*$, similar $k \cdot p$ parameters fit the measured dispersions for GaAs and GaAs:Si (Fig.~\ref{fig:Kink_Full} (b), (d)),
 %the fitting parameters for GaAs and GaAs:Si are similar, suggesting minimal Si-dopant-induced changes to the GaAs band structure.
 %the $k \dot p$ fits are in good agreement with the experimental dispersions of both GaAs and GaAs:Si, demonstrating that our $k \dot p$ model can effectively describe the band structure for small $k$. Notably, the fitting parameters for GaAs and GaAs:Si are similar, suggesting minimal Si-dopant-induced changes to the GaAs band structure.
with minor differences in the fitting parameters likely due to variations in the effective masses.
 %near $\Gamma$, including $\Delta$\textsubscript{SO}, is minimally affected by the incorporation of Si.
Interestingly, at $k ~\approx~0.175$\ Å$^{-1}$, a slight kink in the LH band is apparent in GaAs:Si, but not in GaAs (Fig.~\ref{fig:Kink_Full}).  For $k$-values that exceed $k_{kink}$, the corresponding group velocities, \textit{v}\textsubscript{g}, increase by 30\% (GaAs) and 43\% (GaAs:Si). Since $k_{kink}$ \textgreater $k^*$, further computational methods beyond $k\cdot p$ are needed to inform the origins of the kink. 

Next, we consider the effect of Bi alloying on the GaAs valence band dispersion near $\Gamma$ (Fig.~\ref{fig:Gaps}). The measured band dispersion for GaAs\textsubscript{1-x}Bi\textsubscript{x}:Si is similar to that of both GaAs and GaAs:Si, with easily identified HH, LH, and SO bands (Figs.~\ref{fig:Full}(k)-(o)). However, the signal-to-noise ratio is diminished, likely due to electronic disorder induced by Bi substitutions for Ga. Although the effective masses of GaAs\textsubscript{1-x}Bi\textsubscript{x}:Si are similar to those of GaAs, they are smaller than those of GaAs:Si. Most importantly, a comparison of the relative band positions in GaAs\textsubscript{1-x}Bi\textsubscript{x}:Si vs. GaAs:Si reveals significant differences (Fig.~\ref{fig:Gaps}). Specifically, for GaAs\textsubscript{1-x}Bi\textsubscript{x}:Si, the LH/HH bands reach a VBM at 1.12\,eV below \textit{E}\textsubscript{F}, while the maximum of the SO band is at 1.71\,eV below \textit{E}\textsubscript{F}. Thus, Bi induces an 230\,meV upward shift of the LH/HH bands and a 30\,meV downward shift of the SO band with respect to those in GaAs:Si. In essence, Bi induces a 260\,meV increase in $\Delta$\textsubscript{SO}, from 270\,meV for GaAs:Si to 590\,meV for GaAs\textsubscript{1-x}Bi\textsubscript{x}:Si (Table~\ref{table:Energies}). 
 %\textcolor{orange}{This enhanced $\Delta$\textsubscript{SO} is primarily driven by the upward shift in the valence band which results from hybridization of the GaAs valence band with localized states originating from Bi~\cite{LIU2021}.}
The observed evolution of the HH, LH, and SO bands, as well as the related increase in $\Delta$\textsubscript{SO}, is consistent with band structure predictions for Bi concentrations of ~6\% ~\cite{HONOLKA2019,LIU2021}.

 %\color{red}
 %Comparison of effective masses could go here
 %\color{black}
 %\textcolor{green}{Because of the full $E$-$k$ resolution provided by ARPES, the effective mass of each band can also be determined. The effective mass for each band for the three separate samples are listed in Table~\ref{table:EffectiveMasses}. For GaAs, $m^*_{LH}$ is in good agreement with literature values while our measured $m^*_{HH}$ is roughly twice as large and $m^*_{SO}$ is half as large as literature values. Additionally, the effective masses for each band are slightly enhanced in GaAs:Si compared to GaAs while the values for GaAs\textsubscript{1-x}Bi\textsubscript{x}:Si are closer to those for GaAs.}

 %\begin{table}[t]
 %\centering
 %\setlength{\tabcolsep}{8pt}
 %\caption{Effective masses $m^*$ of the HH, LH, and SO bands in GaAs, GaAs:Si, and GaAs\textsubscript{1-x}Bi\textsubscript{x}:Si along $\Gamma$-X.}
 %\resizebox{\columnwidth}{!}{
 %\begin{tabular}{ c  c  c  c}
 %\hline \hline
 %% Material & $m^*_{\mathrm{HH}}$ ($m_e$) & $m^*_{\mathrm{SO}}$ ($m_e$)\\
 % \hline
 % GaAs & 0.340 &  0.070 \\
 % GaAs:Si & 0.346 & 0.065 \\
 % GaAs\textsubscript{1-x}Bi\textsubscript{x}:Si & 0.385 & 0.078 \\
 % \hline \hline
 %\end{tabular}
 %}
 %\label{table:EffectiveMasses_GX}
 %\end{table}

To explore the origins of the Bi-enhanced band splitting, Si-doped and Bi-alloyed GaAs lattices were modeled with all-electron x2c-DFT calculations. In Fig.\ref{fig:SO-LH_split}, we present a comparison of the calculated $\Delta$\textsubscript{SO} for Ga$_8$As$_8$, Ga$_7$SiAs$_8$, and Ga$_8$As$_7$Bi with the ARPES measured values. 
In both cases, the LH/HH and SO energies consist of multiple degenerate levels. 
Although the simulated doping and alloying concentrations (12.5\%) are higher than the real film concentrations, the Bi-induced enhancement of $\Delta$\textsubscript{SO} is qualitatively consistent with the observed trends. 

 %The valence band structure of GaAs$_{1-x}$Bi$_x$ has been investigated in multiple theoretical studies to probe the origin of increased $\Delta$\textsubscript{SO} in GaAs$_{1-x}$Bi$_x$ alloys. 
 %It has been predicted, using a DFT-based charge patching method, that the increased splitting of the LH/HH and SO bands is due to the relativistic SOC effect of heavier and larger Bi atoms. \cite{zhang2005}.
 %Others have interpreted the change in $\Delta$\textsubscript{SO} mainly as the result of the VBM shifting upwards, mainly due to mixed Bi resonant states, using parameterized valence band anti-crossing (VBAC) and tight binding (TB) models. \cite{usman2011,kudrawiec2012,virkkala2013,joshya2014,bannow2016}

We now consider the primary origins of the increased $\Delta$\textsubscript{SO}, namely, the relative roles of upward shifting of the VBM by mixed Bi resonant states \cite{usman2011,kudrawiec2012,virkkala2013,joshya2014,bannow2016} vs. Bi-induced relativistic SOC\cite{zhang2005}. In our calculations, Bi induces an increase in the CBM - $E_\mathrm{SO}$ energy splitting from 0.78 eV (Ga$_8$As$_8$) to 0.85 eV (Ga$_8$As$_7$Bi); meanwhile, Bi induces a decrease in the CBM - $E_\mathrm{LH/HH}$ energy splitting (i.e. the bandgap energy) from 0.46 eV (Ga$_8$As$_8$) to 0.28 eV (Ga$_8$As$_7$Bi). 
Due to the opposing trends in Bi-induced changes in the CBM - $E_\mathrm{SO}$ and CBM - $E_\mathrm{LH/HH}$ energy splittings, the Bi-induced enhancements of the SO - LH/HH band energy splitting, i.e. $\Delta$\textsubscript{SO}, is attributed primarily to a Bi-induced upward shift of the LH/HH bands, with a minor contribution from Bi-induced enhancement of the relativistic spin-orbit split coupling. 
 %Comparing the experimental band structures between GaAs:Si and GaAs$_{1-x}$Bi$_x$:Si in FIG. \ref{fig:Gaps}, it is also visible that the larger $\Delta_\mathrm{SO}$ is mainly contributed by $E_\mathrm{LH/HH}$ shifting closer to the Fermi level, while $E_\mathrm{SO}$ only changed from 1.68 to 1.71 eV.  
We also discuss the impact of Si doping on the GaAs VB dispersion. Although Si doping lowers the values of $E_\mathrm{SO}$, and $E_\mathrm{LH/HH}$, its effect on the computed and measured band splittings is minimal. Thus Si doping appears to have a negligible effect on $\Delta_\mathrm{SO}$.

Finally, we examine the nature of the x2c-DFT-computed molecular orbital (MO) coefficients. At $\Gamma$, the atomic orbitals (AO) for the LH/HH and SO bands consist primarily of As $p$ orbitals. For Ga$_8$As$_7$Bi, the As $p$ orbital contributions to the valence bands were partially substituted by Bi $p$ orbitals with similar spin alignment, as discussed in the Supplemental Material. This orbital alignment is consistent with previous experimental and theoretical reports \cite{HONOLKA2019,bannow2016}. 
In FIG.~\ref{fig:PDOS}, we present the atomic contributions to the projected density of states (PDOS) in the vicinity of the band gap. 
In the -1 to 0 eV energy range, mixtures of Bi $p$ states and GaAs valence states comprise the observed $E_\mathrm{LH/HH}$ and $E_\mathrm{SO}$. 
Thus, the fully explicit relativistic calculations indicate that enhanced spin-orbit splitting  $\Delta_\mathrm{SO}$ in GaAs$_{1-x}$Bi${_x}$ is primarily due to the upward shift of the VBM due to resonant states introduced by Bi.

 % The Si mixing, while drastically changing the energy of CBM, $E_\mathrm{SO}$, and $E_\mathrm{LH/HH}$, did not affect the band splitting significantly in theory or experiments, suggesting the Si inclusion does not contribute to the change to $\Delta_\mathrm{SO}$.

 % \postit{Some AO labels need to be fixed, see Supplemental Info. I will add a short paragraph here explaining why Bi introduces changes in (i) band gap and (ii) band splitting. The reasoning would be similar to PRB 71, 155201 (2005) and PRB 88, 235201 (2013). \hfill MW}

 % \begin{figure}[ht]
 %     \centering
 %     \includegraphics[width=\linewidth]{exp_SO_band_split.pdf}
 %     \caption{The energy levels of SO and LH bands at $\Gamma$ (a) computed from x2c-DFT ($x=0.125$), and (b) experimentally measured from ARPES as in FIG. \ref{fig:Gaps} ($x=0.06$). $\Delta_{\mathrm{SO}}$ is enhanced by Bi mixing in both calculations and experiments.} 
 %     \label{fig:SO-LH_split}
 % \end{figure}

\begin{table}[t]
\centering
\setlength{\tabcolsep}{8pt}
\caption{Energies of the LH/HH and SO band maxima and the corresponding spin-orbit splitting for GaAs, GaAs:Si, and GaAs\textsubscript{1-x}Bi\textsubscript{x}:Si, from ARPES measurements, for $x\,=\,0.060$.}
\begin{tabular}{ c  c   c   c }
\hline \hline
 Material & \textit{E}\textsubscript{LH/HH} (eV) & \textit{E}\textsubscript{SO} (eV) & $\Delta$\textsubscript{SO} (eV) \\
 \hline
 GaAs & -1.06 & -1.38 & 0.32 \\
 GaAs:Si & -1.35 & -1.68 & 0.33 \\
 GaAs\textsubscript{1-x}Bi\textsubscript{x}:Si & -1.12 & -1.71 & 0.59 \\
 \hline \hline
\end{tabular}
\label{table:Energies}
\end{table}

\subsection{Summary and Outlook}
In summary, we examined the effect of Bi alloying on the GaAs valence band structure. 
 %using high-resolution ARPES, $k\cdot p$ model calculations, and DFT. 
Using droplet-free epitaxial GaAs$_{1-x}$Bi$_x$ films synthesized by MBE, we quantified the local chemistries and morphologies via LEAP, XPS, RBS, XRC, and AFM. In addition, we probed the energetic positions of the LH, HH, and SO valence bands in the vicinity of the $\Gamma$ point using high-resolution ARPES in conjunction with $k\cdot p$ theory and x2C-DFT.  To isolate the effects of Bi alloying on the GaAs VB structure, we consider the measured and computed valence band dispersions for GaAs, GaAs:Si, and GaAs$_{1-x}$Bi$_{x}$:Si films.  
In the vicinity of the GaAs $\Gamma$ point, Si doping induces a rigid shift in VB energies, with a more sudden change in group velocity near $k=0.175$\,Å$^{-1}$.
Furthermore, Bi alloying up to $x_{\mathrm{Bi}}$ = 0.060 induces a 0.23 eV upward shift of the LH maximum, with a corresponding increase in $\Delta_{\mathrm{SO}}$. 
 %In addition, ARPES was used to determine the effective mass in the LH, HH, and SO bands. 
 %DFT and $k\cdot p$ models also support that Bi alloying is responsible for increased spin orbit splitting, with DFT showing that adding Bi leads to increased $\Delta_{\mathrm{SO}}$ primarily through upward shifting of the LH and HH bands, with a smaller contribution due to the increased relativistic effect. 
Using $k\cdot p$ calculations of the $\Gamma$-K band dispersion in conjunction with x2C-DFT, we identify the key role of Bi p-orbitals in the upward shift of the GaAs LH and HH bands that yield the Bi-enhanced $\Delta_\mathrm{SO}$.
This work reveals new insight into the electronic structure of GaAs$_{1-x}$Bi$_x$ alloys, providing a necessary step towards further development of topological insulators and semimetals compatible with III-V compound semiconductors.

\subsection{Acknowledgements}
This research was supported by the National Science Foundation (NSF) through the Materials Research Science and Engineering Center at the University of Michigan, Award No. DMR-2309029. We also gratefully acknowledge support from the NSF (Grant No. ECCS 2240388 and DMR 1810280). This material is also based upon work supported by the NSF CAREER grant under Award No. DMR-2337535. This work used resources of the Advanced Light Source, a U.S. Department of Energy (DOE) Office of Science User Facility under Contract No. DE-AC02-05CH11231.

\pagebreak

\end{document}

\maketitle

\setcounter{equation}{0}
\setcounter{figure}{0}
\setcounter{table}{0}
\setcounter{page}{1}
\makeatletter
\renewcommand{\theequation}{S\arabic{equation}}
\renewcommand{\thefigure}{S\arabic{figure}}
\renewcommand{\thetable}{S\arabic{table}}

\newcommand{\postit}[1]{
    \begin{center}
        \fbox{
            \begin{minipage}{0.8\textwidth}
                \textit{#1}
            \end{minipage}
            }
    \end{center}
}

\clearpage

\tableofcontents

\clearpage

\section{Lattice parameter for GaAs supercell }

The GaAs $2\times2\times2$ supercell was built from a pristine GaAs cell. Both Si-doped and Bi-alloyed GaAs were constructed from the GaAs supercell and retained its original lattice parameter. 

The geometry for GaAs was retrieved from Ref~\cite{ku_synthesis_1968}. Here we present the primitive cell lattice matrix as:
$$
\begin{matrix}
7.99560 &  0.00000 &  0.00000 \\
3.99780 &  6.92439 &  0.00000 \\
3.99780 &  2.30813 &  6.52838 \\
\end{matrix}
$$

We used the same structural parameters for GaAs, Ga$_{1-x}$Si$_{x}$Bi, and GaAs$_{1-x}$Bi$_x$ ($x = 0.125$) are list below (unit: Å).

\resizebox{0.95\textwidth}{!}{
$
\begin{matrix}
\mathrm{Ga}   &0.00000&0.00000&0.00000\\
\mathrm{Ga}   &1.99890&1.15407&3.26419\\
\mathrm{Ga}   &1.99890&3.46220&0.00000\\
\mathrm{Ga}   &3.99780&4.61626&3.26419\\
\mathrm{Ga}   &3.99780&0.00000&0.00000\\
\mathrm{Ga}   &5.99670&1.15407&3.26419\\
\mathrm{Ga}   &5.99670&3.46220&0.00000\\
\mathrm{Ga}   &7.99560&4.61626&3.26419\\
\mathrm{As}   &1.99890&1.15407&0.81605\\
\mathrm{As}   &3.99780&2.30813&4.08024\\
\mathrm{As}   &3.99780&4.61626&0.81605\\
\mathrm{As}   &5.99670&5.77033&4.08024\\
\mathrm{As}   &5.99670&1.15407&0.81605\\
\mathrm{As}   &7.99560&2.30813&4.08024\\
\mathrm{As}   &7.99560&4.61626&0.81605\\
\mathrm{As}   &9.99450&5.77033&4.08024\\
\end{matrix}
\left|
\begin{matrix}
\mathrm{Ga}   &0.00000&0.00000&0.00000\\
\mathrm{Ga}   &1.99890&1.15407&3.26419\\
\mathrm{Ga}   &1.99890&3.46220&0.00000\\
\mathrm{Si}   &3.99780&4.61626&3.26419\\
\mathrm{Ga}   &3.99780&0.00000&0.00000\\
\mathrm{Ga}   &5.99670&1.15407&3.26419\\
\mathrm{Ga}   &5.99670&3.46220&0.00000\\
\mathrm{Ga}   &7.99560&4.61626&3.26419\\
\mathrm{As}   &1.99890&1.15407&0.81605\\
\mathrm{As}   &3.99780&2.30813&4.08024\\
\mathrm{As}   &3.99780&4.61626&0.81605\\
\mathrm{As}   &5.99670&5.77033&4.08024\\
\mathrm{As}   &5.99670&1.15407&0.81605\\
\mathrm{As}   &7.99560&2.30813&4.08024\\
\mathrm{As}   &7.99560&4.61626&0.81605\\
\mathrm{As}   &9.99450&5.77033&4.08024\\
\end{matrix}
\right|
\begin{matrix}
\mathrm{Ga}   &0.00000&0.00000&0.00000\\
\mathrm{Ga}   &1.99890&1.15407&3.26419\\
\mathrm{Ga}   &1.99890&3.46220&0.00000\\
\mathrm{Ga}   &3.99780&4.61626&3.26419\\
\mathrm{Ga}   &3.99780&0.00000&0.00000\\
\mathrm{Ga}   &5.99670&1.15407&3.26419\\
\mathrm{Ga}   &5.99670&3.46220&0.00000\\
\mathrm{Ga}   &7.99560&4.61626&3.26419\\
\mathrm{Bi}   &1.99890&1.15407&0.81605\\
\mathrm{As}   &3.99780&2.30813&4.08024\\
\mathrm{As}   &3.99780&4.61626&0.81605\\
\mathrm{As}   &5.99670&5.77033&4.08024\\
\mathrm{As}   &5.99670&1.15407&0.81605\\
\mathrm{As}   &7.99560&2.30813&4.08024\\
\mathrm{As}   &7.99560&4.61626&0.81605\\
\mathrm{As}   &9.99450&5.77033&4.08024\\
\end{matrix}
$
}

\section{Molecular orbitals of valence and conduction bands}
Here we present detailed x2c-DFT results for GaAs and GaAs$_{1-x}$Bi$_x$. In TABLE \ref{table:GaAs MO} and TABLE \ref{table:GaAsBi MO}, we list the molecular orbital (MO) energies and their major contributing atomic orbitals (AO) for the valence and conduction bands. 

The weight of AO is computed with the MO coefficient matrix $C$ and overlap matrix $S$ as
\begin{equation}
    W_{\mu i} = C^{*}_{\mu i} \cdot \sum_{\nu}S_{\mu\nu}C_{\nu i}.
\end{equation}
$W_{\mu i}$ represents the estimated contribution of AO $\mu$ to MO $i$. 
For the host material Ga and As atoms, any AOs with more than 10\% are listed.
For the impurity Bi atoms, any AOs with more than 1\% are listed, considering the much lower concentration of Bi than Ga and As. 

% \usepackage{multirow}
\begin{table}[h!]
\begin{tabular}{cccc}
\hline
Band                & \multicolumn{1}{c}{MO index} & \multicolumn{1}{c}{$E_{\mathrm{MO}} - E_{\mathrm{F}}$ (eV)} & Major AO        \\
\hline
\multirow{3}{*}{Valence SO} & \multirow{3}{*}{506}         & \multirow{3}{*}{-0.549793}     
& As 4px $\alpha$ (0.123) \\
&&& As 4py $\alpha$ (0.123) \\
&&& As 4pz $\beta$  (0.123) \\

\hline
\multirow{3}{*}{Valence SO} & \multirow{3}{*}{507}         & \multirow{3}{*}{-0.549793}     & As 4px $\beta$ (0.123)   \\
&&& As 4py $\beta$ (0.123) \\
&&& As 4pz $\alpha$ (0.123) \\

\hline
\multirow{4}{*}{Valence LH/HH} & \multirow{4}{*}{508}         & \multirow{4}{*}{-0.228866}    
& As 4px $\alpha$ (0.171) \\
&&& As 4py $\alpha$ (0.188)\\
&&& As 5px $\alpha$ (0.141) \\
&&& As 5py $\alpha$ (0.154)\\
\hline
\multirow{4}{*}{Valence LH/HH} & \multirow{4}{*}{509}         & \multirow{4}{*}{-0.228866}     & As 4px $\beta$ (0.171)  \\
&&& As 4py $\beta$ (0.188) \\
&&& As 5px $\beta$ (0.141) \\
&&& As 5py $\beta$ (0.154) \\
\hline
\multirow{2}{*}{Valence LH/HH} & \multirow{2}{*}{510}         & \multirow{2}{*}{-0.228805}     
& As 4pz $\alpha$ (0.224) \\
&&& As 5pz $\alpha$ (0.185) \\
\hline
\multirow{2}{*}{Valence LH/HH} & \multirow{2}{*}{511}         & \multirow{2}{*}{-0.228805}
& As 4pz $\beta$ (0.224) \\
&&& As 5pz $\beta$ (0.185) \\
\hline
\multirow{3}{*}{Conduction} &  \multirow{3}{*}{512}  &  \multirow{3}{*}{0.228805} &  Ga 4s $\alpha$ (0.466) \\
 &    &    &  Ga 5s $\alpha$ (0.410) \\
 &    &    &  As 4s $\alpha$ (0.307) \\
\hline
\multirow{3}{*}{Conduction} &  \multirow{3}{*}{513}  &  \multirow{3}{*}{0.228805} &  Ga 4s $\beta$ (0.466) \\
 &    &    &  Ga 5s $\beta$ (0.410) \\
 &    &    &  As 4s $\beta$ (0.307) \\
\hline
\end{tabular}
\caption{x2c-DFT results for GaAs at $\Gamma$. Selected AOs to each energy level are listed based on their MO eigenvectors. }
\label{table:GaAs MO}
\end{table}

\begin{table}
\begin{tabular}{cccc}
\hline
Band                & \multicolumn{1}{c}{MO index} & \multicolumn{1}{c}{$E_{\mathrm{MO}} - E_{\mathrm{F}}$ (eV)} & Major AO contribution       \\
\hline
\multirow{3}{*}{Valence SO} & \multirow{3}{*}{556} & \multirow{3}{*}{-0.713126} 
& As 4px $\alpha$ (0.122), Bi 7px $\alpha$ (0.013)\\
&&& As 4py $\alpha$ (0.122), Bi 7py $\alpha$ (0.013)\\
&&& As 4pz $\beta$ (0.121), Bi 7pz $\beta$ (0.013) \\
\hline
\multirow{3}{*}{Valence SO} & \multirow{3}{*}{557} & \multirow{3}{*}{-0.713126} 
& As 4px $\beta$ (0.122), Bi 7px $\beta$ (0.013)\\
&&& As 4py $\beta$ (0.122), Bi 7py $\beta$ (0.013)\\
&&& As 4pz $\alpha$ (0.121), Bi 7pz $\alpha$ (0.013) \\
\hline
\multirow{4}{*}{Valence LH/HH} & \multirow{4}{*}{558}         & \multirow{4}{*}{-0.139898}     
& As 4px $\beta$ (0.146), Bi 7px $\beta$ (0.028) \\
&&& As 4py $\beta$ (0.132), Bi 7py $\beta$ (0.025) \\
&&& As 5px $\beta$ (0.125), Bi 8px $\beta$ (0.020) \\
&&& As 5py $\beta$ (0.112), Bi 8py $\beta$ (0.019) \\
\hline
\multirow{4}{*}{Valence LH/HH} & \multirow{4}{*}{559}         & \multirow{4}{*}{-0.139898}     
& As 4px $\alpha$ (0.146), Bi 7px $\alpha$ (0.028) \\
&&& As 4py $\alpha$ (0.132), Bi 7py $\alpha$ (0.025) \\
&&& As 5px $\alpha$ (0.125), Bi 8px $\alpha$ (0.020) \\
&&& As 5py $\alpha$ (0.112), Bi 8py $\alpha$ (0.019) \\
\hline
\multirow{5}{*}{Valence LH/HH} & \multirow{5}{*}{560}         & \multirow{5}{*}{-0.139817}     
& As 4pz $\alpha$ (0.125), Bi 7pz $\alpha$ (0.024) \\
&&& As 5pz $\alpha$ (0.106), Bi 8pz $\alpha$ (0.018) \\
&&& Bi 7px $\beta$ (0.013) \\
&&& Bi 7pz $\beta$ (0.012) \\ 
&&& Bi 7py $\alpha$ (0.012) \\
\hline
\multirow{5}{*}{Valence LH/HH} & \multirow{5}{*}{561}         & \multirow{5}{*}{-0.139817}     
& As 4pz $\beta$ (0.125), Bi 7pz $\beta$ (0.024) \\
&&& As 5pz $\beta$ (0.106), Bi 8pz $\beta$ (0.018) \\
&&& Bi 7px $\alpha$ (0.013) \\
&&& Bi 7pz $\alpha$ (0.012) \\ 
&&& Bi 7py $\beta$ (0.012) \\
\hline
\multirow{2}{*}{Conduction} &  \multirow{2}{*}{562}  &  \multirow{2}{*}{0.139817}  
&  Ga 4s $\alpha$ (0.456), Ga 5s $\alpha$ (0.392) \\
&    &    & As 4s $\alpha$ (0.279), Bi 6s $\alpha$ (0.012) \\
\hline
\multirow{2}{*}{Conduction} &  \multirow{2}{*}{563}  &  \multirow{2}{*}{0.139817}  
&  Ga 4s $\beta$ (0.456), Ga 5s $\beta$ (0.392) \\
&    &    & As 4s $\beta$ (0.279), Bi 6s $\beta$ (0.012) \\
\hline
\end{tabular}
\caption{x2c-DFT results for GaAs$_{1-x}$Bi$_x$ ($x=0.125$) at $\Gamma$. Selected AOs to each energy level are listed based on their MO eigenvectors. }
\label{table:GaAsBi MO}
\end{table}

% \postit{The AO contribution needs to be fixed. I used the squared MO coefficient matrix $w = C^2$, which is not fully correct. The overlap matrix S should be involved, so the contribution considers occupation: $w = \sum CSC$. \hfill MW}

Note that in Table \ref{table:GaAsBi MO}, the Bi 7$p$ orbitals predominantly participated in the valance bands at $\Gamma$, instead of the valence shell 6$p$ orbitals. 
This might be attributed to: 
(i) the full electron basis (x2c-SVPall) is over complete for this supercell x2c-DFT calculation, leading to some liner dependency; 
(ii) the GaAs$_{1-x}$Bi$_x$ ($x=0.125$) calculation used the same lattice parameter as pristine GaAs. For heavy Bi atoms, this implies a strained geometry.

\section{Effective Masses}

The effective masses of each band along the $\Gamma$-K and $\Gamma$-X directions are enumerated in Table~\ref{table:EffectiveMasses}. Each band was parabolically fit around $\Gamma$ and the fit equations were then used to compute $\frac{d^2E}{dk^2}$ to then find the effective mass $m^* = \frac{\hbar^2}{\frac{d^2E}{dk^2}}$. Due to experimental broadening, the effective mass of the LH band could not be found for all samples along $\Gamma$-X and for GaAs\textsubscript{1-x}Bi\textsubscript{x}:Si along $\Gamma$-K.

\begin{table}[h]
%\centering
%\setlength{\tabcolsep}{8pt}
\caption{Effective masses $m^*$ of the HH, LH, and SO bands in GaAs, GaAs:Si, and GaAs\textsubscript{1-x}Bi\textsubscript{x}:Si. along $\Gamma$-K and $\Gamma$-X. Those marked with a '-' could not be measured due to experimental broadening.}
%\resizebox{\columnwidth}{!}{
\begin{tabular}{ c  c  c  c  c  c  c}
\hline \hline
\multirow{2}{4em}{Material} & \multicolumn{2}{c}{$m^*_{\mathrm{HH}}$ ($m_e$)} & \multicolumn{2}{c}{$m^*_{\mathrm{LH}}$ ($m_e$)} & \multicolumn{2}{c}{$m^*_{\mathrm{SO}}$ ($m_e$)} \\

  & $\Gamma$-K & $\Gamma$-X & $\Gamma$-K & $\Gamma$-X & $\Gamma$-K & $\Gamma$-X \\
 \hline
 GaAs & 0.959 & 0.340 & 0.095 & - & 0.060 & 0.070 \\
 GaAs:Si & 1.127 & 0.346 & 0.102 & - & 0.070 & 0.065 \\
 GaAs\textsubscript{1-x}Bi\textsubscript{x}:Si & 1.011 & 0.385 & - & - & 0.061 & 0.078 \\
 \hline \hline
\end{tabular}
%}
\label{table:EffectiveMasses}
\end{table}

\section{$k\cdot p$ Calculation}
In the absence of spin-orbit coupling, the three hole bands at the $\Gamma$ point form a triplet (three-fold degeneracy), corresponding to the $\Gamma_5$ irreducible representation of the point group $T_d$. With this symmetry information, we can write down the $k \cdot p$ Hamiltonian for these three hole bands near the $\Gamma$ point. Including spin degrees of freedom, the Hamiltonian becomes a $6 \times 6$ matrix. Since we are interested in the vicinity of the $\Gamma$ point (i.e., small wavevector), the Hamiltonian can be expanded as a power series in $\mathbf{k}$, retaining only the leading-order terms. Below, we list all symmetry-allowed terms at leading order:
\begin{align}
H_0=&\alpha_{\Gamma_1} (k_x^2 + k_y^2 + k_z^2)
\begin{pmatrix}
1 & 0 & 0 & 0 & 0 & 0 \\
0 & 1 & 0 & 0 & 0 & 0 \\
0 & 0 & 1 & 0 & 0 & 0 \\
0 & 0 & 0 & 1 & 0 & 0 \\
0 & 0 & 0 & 0 & 1 & 0 \\
0 & 0 & 0 & 0 & 0 & 1 \\
\end{pmatrix}
+\,\alpha_{\Gamma_5} 
\begin{pmatrix}
0 & 0 & k_x k_y & 0 & k_x k_z & 0 \\
0 & 0 & 0 & k_x k_y & 0 & k_x k_z \\
k_x k_y & 0 & 0 & 0 & k_y k_z & 0 \\
0 & k_x k_y & 0 & 0 & 0 & k_y k_z \\
k_x k_z & 0 & k_y k_z & 0 & 0 & 0 \\
0 & k_x k_z & 0 & k_y k_z & 0 & 0 \\
\end{pmatrix}\nonumber \\[2ex]
+\, &\alpha_{\Gamma_3} \frac{-2 k_z^2 + k_x^2 + k_y^2}{\sqrt{6}}
\begin{pmatrix}
1 & 0 & 0 & 0 & 0 & 0 \\
0 & 1 & 0 & 0 & 0 & 0 \\
0 & 0 & 1 & 0 & 0 & 0 \\
0 & 0 & 0 & 1 & 0 & 0 \\
0 & 0 & 0 & 0 & -2 & 0 \\
0 & 0 & 0 & 0 & 0 & -2 \\
\end{pmatrix}
+\alpha_{\Gamma_3} \frac{\sqrt{3}}{\sqrt{2}} (k_x^2 - k_y^2)
\begin{pmatrix}
1 & 0 & 0 & 0 & 0 & 0 \\
0 & 1 & 0 & 0 & 0 & 0 \\
0 & 0 & -1 & 0 & 0 & 0 \\
0 & 0 & 0 & -1 & 0 & 0 \\
0 & 0 & 0 & 0 & 0 & 0 \\
0 & 0 & 0 & 0 & 0 & 0 \\
\end{pmatrix} 
\end{align}
where $\alpha_{\Gamma_1}$, $\alpha_{\Gamma_3}$, and $\alpha_{\Gamma_5}$ are three control parameters.

The spin-orbit (SO) coupling introduces an additional term to the $k \cdot p$ Hamiltonian:
\begin{align}
H_{SO} = \frac{\Delta_{SO}}{3} \begin{pmatrix}
-1 & 0 & i & 0 & 0 & -1 \\
0 & -1 & 0 & -i & 1 & 0 \\
-i & 0 & -1 & 0 & 0 & i \\
0 & i & 0 & -1 & i & 0 \\
0 & 1 & 0 & -i & -1 & 0 \\
-1 & 0 & -i & 0 & 0 & -1
\end{pmatrix}
\end{align}
where $\Delta_{SO}$ is the strength of the spin-orbit coupling. The SO coupling lifts the degeneracy of the three hole bands, splitting them into a two-fold degenerate $\Gamma_7$ doublet and a four-fold degenerate $\Gamma_8$ quartet. Here, we set the energy of the quartet as the reference point ($E = 0$) by introducing appropriate diagonal components to $H_{SO}$. It should be emphasized that $E=0$ is simply a reference, and its absolute value can be shifted.

By computing the eigenvalues of the $H_0+H_{SO}$, we obtain the dispersion relation near the Gamma point. Along the $\Gamma-X$ direction, the dispersion relation is
\begin{align}
E_1 &= \frac{1}{3} k^2 \left(-3\alpha_{\Gamma_1} + \sqrt{6}\,\alpha_{\Gamma_3} \right)+\epsilon_0 \\[1ex]
E_2 &= \frac{1}{6} \left( -k^2 \left( 6\alpha_{\Gamma_1} + \sqrt{6}\,\alpha_{\Gamma_3} \right) - 3\Delta_{SO} + \sqrt{54 k^4 \alpha_{\Gamma_3}^2 - 6\sqrt{6}\,k^2 \alpha_{\Gamma_3} \Delta_{SO} + 9\Delta_{SO}^2} \right)+\epsilon_0\\[1ex]
E_3 &= \frac{1}{6} \left( -k^2 \left( 6\alpha_{\Gamma_1} + \sqrt{6}\,\alpha_{\Gamma_3} \right) - 3\Delta_{SO} - \sqrt{54 k^4 \alpha_{\Gamma_3}^2 - 6\sqrt{6}\,k^2 \alpha_{\Gamma_3} \Delta_{SO} + 9\Delta_{SO}^2} \right) +\epsilon_0
\end{align}
Each of these three bands is two-fold degenerate due to spin degrees of freedom. Additionally, $\epsilon_0$ has been added as a constant offset so as to compare directly to experiment. Here $\epsilon_3$ is the dispersion of the doublet band, and $\epsilon_1$ and $\epsilon_2$ are the dispersions of the two bands that form the quartet at the $\Gamma$ point. And it is easy to check that at $\Gamma$ point, we have $E_1=E_2=+\epsilon_0$ and $E_3=\epsilon_0-\Delta_{SO}$.

Along the $\Gamma-K$ direction, the dispersion is 
\begin{align}
\epsilon_1 &=
\frac{1}{6} \Bigg[
    -2 \left(3 k^2 \alpha_{\Gamma_1} + \Delta_{SO}\right) 
%    \notag\\&
    \quad + \frac{
        3 k^4 \left(2 \alpha_{\Gamma_3}^2 + \alpha_{\Gamma_5}^2\right) + 4 \Delta_{SO}^2
    }{
        \left(
            6 \sqrt{6}\, k^6 \alpha_{\Gamma_3}^3
            - 9 \sqrt{6}\, k^6 \alpha_{\Gamma_3} \alpha_{\Gamma_5}^2
            - 8 \Delta_{SO}^3
            + D
        \right)^{1/3}
    } \notag\\
    &\quad + \left(
            6 \sqrt{6}\, k^6 \alpha_{\Gamma_3}^3
            - 9 \sqrt{6}\, k^6 \alpha_{\Gamma_3} \alpha_{\Gamma_5}^2
            - 8 \Delta_{SO}^3
            + D
        \right)^{1/3}
\Bigg] + \epsilon_0
\end{align}

\begin{align}
\epsilon_2 &=
\frac{1}{24} \Bigg[
    -8 \left(3 k^2 \alpha_{\Gamma_1} + \Delta_{SO}\right)\quad + \frac{
        2i(i + \sqrt{3})\left[3 k^4 \left(2 \alpha_{\Gamma_3}^2 + \alpha_{\Gamma_5}^2\right) + 4 \Delta_{SO}^2\right]
    }{
        \left(
            6 \sqrt{6}\, k^6 \alpha_{\Gamma_3}^3
            - 9 \sqrt{6}\, k^6 \alpha_{\Gamma_3} \alpha_{\Gamma_5}^2
            - 8 \Delta_{SO}^3
            + D
        \right)^{1/3}
    } \notag\\
    &\quad - 2(1+i\sqrt{3}) \left(
            6 \sqrt{6}\, k^6 \alpha_{\Gamma_3}^3
            - 9 \sqrt{6}\, k^6 \alpha_{\Gamma_3} \alpha_{\Gamma_5}^2
            - 8 \Delta_{SO}^3
            + D
        \right)^{1/3}
\Bigg] + \epsilon_0
\end{align}

\begin{align}
\epsilon_3 &=
\frac{1}{24} \Bigg[
    -8 \left(3 k^2 \alpha_{\Gamma_1} + \Delta_{SO}\right) 
    \quad - \frac{
        2i(-i + \sqrt{3})\left[3 k^4 \left(2 \alpha_{\Gamma_3}^2 + \alpha_{\Gamma_5}^2\right) + 4 \Delta_{SO}^2\right]
    }{
        \left(
            6 \sqrt{6}\, k^6 \alpha_{\Gamma_3}^3
            - 9 \sqrt{6}\, k^6 \alpha_{\Gamma_3} \alpha_{\Gamma_5}^2
            - 8 \Delta_{SO}^3
            + D
        \right)^{1/3}
    } \notag\\
    &\quad + 2i(i+\sqrt{3}) \left(
            6 \sqrt{6}\, k^6 \alpha_{\Gamma_3}^3
            - 9 \sqrt{6}\, k^6 \alpha_{\Gamma_3} \alpha_{\Gamma_5}^2
            - 8 \Delta_{SO}^3
            + D
        \right)^{1/3}
\Bigg] + \epsilon_0
\end{align}

where we defined quantity $D$ as
\begin{align}
D = \sqrt{
  -\left[3 k^4 \left(2 \alpha_{\Gamma_3}^2 + \alpha_{\Gamma_5}^2\right) + 4 \Delta_{SO}^2\right]^3
  + \left[3\sqrt{6}\,k^6 \alpha_{\Gamma_3} (2\alpha_{\Gamma_3}^2 - 3\alpha_{\Gamma_5}^2) - 8\Delta_{SO}^3\right]^2
}.
\end{align}

The variables $\alpha_{\Gamma_1}$, $\alpha_{\Gamma_3}$, $\alpha_{\Gamma_5}$, $\Delta_{SO}$, and $\epsilon_0$ can then be determined by comparing to our measured ARPES data. First, $\Delta_{SO}$ and $\epsilon_0$ are fixed to the values measured by ARPES. $\alpha_{\Gamma_1}$, $\alpha_{\Gamma_3}$, and $\alpha_{\Gamma_5}$ are left as fitting parameters and are determined by fitting the equation for the LH band to the measured LH dispersion in the range $-0.125\text{\AA}^{-1}<k<0.125\text{\AA}^{-1}$. For GaAs, the fitting parameters are: $\alpha_{\Gamma_1}=36.176$, $\alpha_{\Gamma_3}=26.469$, and $\alpha_{\Gamma_5}=76.928$. For GaAs:Si, the fitting parameters are: $\alpha_{\Gamma_1}=30.862$, $\alpha_{\Gamma_3}=26.368$, and $\alpha_{\Gamma_5}=64.5876$.

The $k\cdot p$ Hamiltonian used is only valid for small $k$. To determine an approximate region for which the dispersion equations are valid ($k<k^*$), we define 
\begin{align}
\epsilon = \epsilon_1 + \epsilon_2 + \epsilon_3.
\end{align}
We then fit $\epsilon$ using a polynomial $y$ in even powers of $k$ as
\begin{align}
y = a + bk^2+ck^4+dk^6+ek^8.
\end{align}
We define $k*$ as the value that satisfies $b(k^*)^2=c(k^*)^4$. We thus assume that our $k\cdot p$ model holds for all $|k|<k^*$. For GaAs, $k^*=0.149\text{\AA}^{-1}$. For GaAs:Si, $k^*=0.165\text{\AA}^{-1}$

\section{Surface charging effects}
For the GaAs\textsubscript{1-x}Bi\textsubscript{x}:Si layers, ARPES and XPS measurements were performed on two pieces of the MBE-grown sample, each with equivalent compositions. For piece 2, all core level emissions were shifted to higher binding energies in comparison with those of piece 1.  For example, a 90 meV difference in the binding energies for the Bi 4\textit{f} core level is apparent, presumably due to electrostatic charging of piece 2. Thus, for piece 2, all spectra are shifted upwards by 90m eV to match the Bi 4\textit{f} core levels for piece 1. The resulting energies of the heavy hole/light hole bands at $\Gamma$ are consistent between the two pieces, while the energy of the split-off band at $\Gamma$ differs by only 10meV. %This shows the homogeneity of the films and demonstrates the repeatability of the measured value of $\Delta_\mathrm{SO}$. This agreement after a rigid shift in energy suggests that Piece 2 was indeed charged relative to Piece 1 and thus the 90meV shifted energies for Piece 2 are used throughout this manuscript. 
%We note that while we are not able to definitively confirm that Piece 1, or the GaAs or GaAs:Si pieces, were not charged themselves, 
Since any charging induces a rigid shift in binding energies, the relative differences in band positions (e.g. $\Delta_{\text{SO}}$) remain unaffected. Therefore, our conclusions regarding the effects of Bi alloying on $\Delta_{\text{SO}}$ remain regardless of the possibility of charging in these semiconducting samples.

%\bibliography{reference}